\documentstyle[preprint,aps]{revtex}
\tightenlines
\def\journal#1, #2, #3#4, #5#6#7#8    {
    {#1~}{\bf #2}, #3#4 (#5#6#7#8)}

\def\pra{\journal Phys. Rev. A, }

\def\prl{\journal Phys. Rev. Lett., }

\def\jp{\journal J. Phys. A, }

\newcommand{\beq}[1]{\begin{equation}\label{#1}}
\newcommand\eeq{\end{equation}}
\newcommand{\ba}[1]{\begin{eqnarray}\label{#1}}
\newcommand{\baa}{\begin{eqnarray}}
\newcommand\ea{\end{eqnarray}}
\newcommand{\bee}{\begin{equation}}
\def\nn{\nonumber \\}
\def\l{\lambda}

\def\f{\varphi}

\def\e{\tilde e}

\newcommand{\h}{Hamiltonian}
\newcommand{\an}{angular momentum}
\newcommand{\B}[1]{{\bf #1}}
\def\hlf{\frac{1}{2}}

\newcommand{\lsim}{\stackrel{\rm <} {\scriptstyle \sim}}

\begin{document}
%\input{hrvzn}
%\draft
%{\Huge{\bf
\title{Perturbative Spectrum of
 Trapped Weakly Interacting Bosons in Two Dimensions}
\author{Velimir Bardek, Larisa Jonke, and Stjepan Meljanac
\footnote{e-mail address: 
bardek@thphys.irb.hr \\ \hspace*{3cm} larisa@thphys.irb.hr \\ 
\hspace*{3cm} meljanac@thphys.irb.hr}} 
\address{Theoretical Physics Division,\\
Rudjer Bo\v skovi\'c Institute, P.O. Box 180,\\
HR-10002 Zagreb, CROATIA}
\maketitle
\begin{abstract}
We study a trapped Bose-Einstein condensate under rotation in the limit 
of weak, translational and rotational invariant two-particle interactions.
We use the perturbation-theory approach (the large-$N$ expansion) to 
calculate the ground-state energy and the excitation spectrum in the 
asymptotic limit where the total number of particles $N$ goes to 
infinity while keeping the total \an \ $L$ finite. Calculating the 
probabilities of different configurations of angular momentum
  in the exact eigenstates
gives us a clear view of the physical content of excitations. We briefly discuss
the case of repulsive contact interaction.
\end{abstract}
\vspace{1cm}
\hspace{1.7cm} PACS number(s): 03.75.Fi, 05.30.Jp, 03.65.Fd, 67.40Db

\newpage
The study of low-lying excitations of the weakly interacting, trapped 
Bose-Einstein condensed gas under rotation
 is of considerable experimental\cite{10,11} and
theoretical interest\cite{6}. Theoretical studies have focused  on the 
Thomas-Fermi limit of strong interactions\cite{6}, as well as on the 
limit of weak interactions\cite{5,mott,9,7}, which we consider in this paper.
Wilkin et al.\cite{5} studied the case of attractive interaction, 
and Mottelson
 and Kavulakis et al.\cite{mott} developed a theory for repulsive 
interactions. They compared the mean-field approach 
and exact numerical results obtained 
by diagonalization in a subspace of degenerate states\cite{9}. Bertsch and 
Papenbrock\cite{7} 
performed numerical diagonalization for small systems 
and showed that the interaction energy of the 
lowest-energy states decreases linearly with \an \ $L$. 
Nakajima and Ueda\cite{18} found through an extensive numerical study, in the 
limit where the \an \ per particle is much smaller than one, that 
low-lying excitation energies, measured from the energy of the lowest state are
given by $0.795\,n(n-1)$, where $n$ is the number of octupole excitations.
Recently, Kavoulakis et al.\cite{22} rederived these results analytically
with use of the diagrammatic perturbation-theory approach
in the asymptotic limit 
$N\rightarrow\infty$. 
In this paper we present a systematic method for calculating the excitation 
spectrum for the  weak, translationally and rotationally 
symmetric interaction  in the 
asymptotic limit, where the total number of particles $N$ goes to
infinity, while keeping the total \an \ $L$ finite. We also discuss the
probabilities of different configurations of the angular momentum
in the exact eigenstates.

Our starting point is the two-dimensional \h \ $H=H_0+V$, where
\beq 1
H_0=\sum_{i=1}^N\left(-\hlf{\B\nabla_i}^2+\hlf\B r_i^2\right)\eeq
is the one-particle part, including the kinetic energy of the particles, and 
the potential energy due to the trapping potential. The trapping potential is 
approximated by a two-dimensional, isotropic harmonic oscillator with the
frequency set to one. The 
system is in the ground state for the motion in the direction of the axis of 
rotation.
The two-body interaction between the particles is given by
\bee\label{1a} V=\sum_{i<j}^Nv(|\B r_i-\B r_j|),\eeq
where an arbitrary potential $v$ possesses translational and rotational 
symmetries. We also assume that the interaction $v$ is weak.
This allows us to work within the subspace of single-particle states with 
no radial excitations
\bee\label{1b}\psi_n(z)=(\pi n!)^{-1/2}z^n\exp(-\hlf|z|^2),\eeq
 where 
$z=x+iy$ and $n$ is the \an \ quantum number. The energy levels and the 
corresponding wave functions are found by diagonalizing the interaction 
$V$ in this Hilbert space. Basis functions for the many-body problem are
$\psi(z_1,z_2,\ldots,z_N)=\f(z_1,z_2,\ldots,z_N)\prod_{i=1}^N
\exp(-\hlf|z_i|^2)$, where $\f$ is a homogeneous polynomial of degree $L$.
For simplicity, we omit the exponentials from the wave functions.
Suitable basis functions for such 
polynomials are given by
\beq 3
B_{\l}(z_1,z_2,\ldots,z_N)=\frac{1}{\nu_1!\nu_2!\cdots \nu_p!}
\sum_{i_1,i_2,\ldots,i_q=1}^N\hspace{-0.6cm}^{'}
z_{i_1}^{\l_1}z_{i_2}^{\l_2}\cdots z_{i_q}^{\l_q},\eeq
where the 
set $\{\l_1,\l_2,\ldots,\l_q\}$ denotes any partition of $L$ such that 
$\sum_{i=1}^q\l_i=L$ and $\l_1\geq\l_2\geq\cdots\geq\l_q>0$ for 
$q\leq N$\cite{16}. 
The prime denotes the sum over mutually different indices 
$i_1,i_2,\ldots,i_q$, 
while the numbers $\nu_1,\nu_2,\ldots,\nu_p$ 
denote the frequencies of appearance 
of equal $\l$'s. 
Note that the number of distinct monomial terms
$z_{i_1}^{\l_1}z_{i_2}^{\l_2}\cdots z_{i_q}^{\l_q}$ in $B_{\l}$ is given by
$d_{\l}=N(N-1)\cdots(N-q+1)/(\nu_1!\nu_2!\cdots \nu_p!)$, 
where $\nu_1+\nu_2+\cdots+\nu_p=q$.
Owing to the translational and rotational symmetries of the two-particle 
interaction $v(|\B r_i-\B r_j|)$, for  non-negative integers $n$ and $m$ we have
\ba 4
&&v(|z_1-z_2|)\,(z_1+z_2)^n(z_1-z_2)^{2m}P(z_3,z_4,\ldots,z_N)\nn
&=&c_{2m}(z_1+z_2)^n(z_1-z_2)^{2m}
P(z_3,z_4,\ldots,z_N), \ea
where $P$ denotes an arbitrary polynomial.
The coefficient $c_n$ is given by
\beq 5
c_n=\frac{\int_0^{\infty}dr\, r^{2n+1}v(r)\exp(-r^2/2)}
{\int_0^{\infty}dr\, r^{2n+1}\exp(-r^2/2)}\; .\eeq 
and represents the interaction energy 
$v(r)$ of the relative motion of two bosons in the single-particle state 
$r^n\exp(-r^2/2)$ with the \an \ $n$.
To proceed, let us now define symmetric functions of two variables
\beq 7
b_{ij}(z_1,z_2)=\hlf(z_1^iz_2^j+z_1^jz_2^i),\;i\geq j.\eeq
The action of the potential $v(|\B r_i-\B r_j|)$ on $b_{ij}$ is given by
\beq 8
v(|z_1-z_2|)b_{ij}(z_1,z_2)=\sum_{l=0}^{[\frac{n}{2}]}\alpha_{ij}^{kl}
b_{kl}(z_1,z_2),\eeq
where $i+j=k+l=n$. This restriction is a consequence of the 
conservation of total \an \ for a rotationally symmetric potential.
Also,  the coefficients $\alpha_{ij}^{kl}$ 
\beq 9
\alpha_{ij}^{kl}=\frac{2-\delta_{kl}}{2^n}\sum_{p=0}^{[\frac{n}{2}]}
c_{2p}S_{i,j}^{2p}S_{n-2p,2p}^l, \eeq
where
\beq a
S_{i,j}^q=\sum_{r+s=q}(-)^s{i\choose r}
{j\choose s} ,\eeq
satisfy the 
summation rule $\sum_{l=0}^{[\frac{n}{2}]}\alpha_{ij}^{kl}=c_0$, 
as a consequence of translational symmetry.
The coefficients $\alpha_{ij}^{kl}$ in fact 
represent the two-body matrix element 
$V_{ijkl}$ (see Ref.\cite{13})
of the interaction potential $V$. 
Next, by using Eqs.(\ref{3}) and (\ref{4}) 
we obtain
\beq b 
V\,B_1^n(z_1,z_2,\ldots,z_N)=c_0{N\choose 2}B_1^n(z_1,z_2,\ldots,z_N)\;
{\rm for}\;B_1=\sum_{i=1}^Nz_i. \eeq
As a result, $B_1^n(z_1,z_2,\ldots,z_N)$ is an exact eigenstate  and the 
corresponding
eigenvalue  is $c_0{N\choose 2}$.
Furthermore, owing to translational invariance 
the action of the potential $V$ on 
the product $B_1^nB_{\l}$ reduces to
\beq c
V\,B_1^nB_{\l}=B_1^nVB_{\l} \eeq
for any $n$ and partition $\l$. Specially, if $A$ is an eigenstate with energy 
$E$, then $B_1^nA$ is also an eigenstate with the same energy.

Generally, for any partition $\l$ we find
\bee\label{n1}
VB_{\l}=\sum_{\mu}a_{\l}^{\mu}B_{\mu},\eeq
where $\mu$ is the partition obtained by substituting a pair of numbers 
$\{k,l\}$ for a
$\{i,j\}$ in partition $\l$, such that $i\geq 
j$, $k\geq l$, and $i+j=k+l$,  for all distinct pairs $\{i,j\}$ and all allowed
 $\{k,l\}$.
Note that any partition $\l=\{\l_1,\cdots,\l_q\}$ can be written as
$\l=0^{n_0}1^{n_1}2^{n_2}\cdots l^{n_l}\cdots$, with
$\sum_{i}n_i(\l)=N$ and $\sum_{i}in_i(\l)=L$. 
In the second quantized approach, the number $n_i(\l)$ can be 
interpreted as  the number
of particles with the \an \ $i$ in the partition $\l$.
The diagonal coefficient is 
\bee\label{n2}
a_{\l}^{\l}=\sum_{\{i,j\}}\alpha_{ij}^{ij}\frac{2-\delta_{i,j}}{2}n_i(\l)
[n_j(\l)-\delta_{i,j}], \eeq
where the sum goes over all distinct  pairs $\{i,j\},\;i\geq j\geq 0$ 
contained in partition $\l$.
The sum contains 
${K_1\choose 2}+K_2$ terms, where $K_1$ is the number of  $n_i$'s  greater 
than zero ,
 and $K_2$ is the number of $n_i$'s greater than one, in partition $\l$ 
 ($i\geq 0$).
The nondiagonal coefficient can be expressed as
\bee\label{n3}
a_{\l}^{\mu}=\alpha_{ij}^{kl}\frac{2-\delta_{i,j}}{2}
n_k(\mu)[n_l(\mu)-\delta_{k,l}],\eeq
where $\{i,j\}$ ($\{k,l\}$) are contained in partition $\l$ 
($\mu$), respectively.
The  general matrix element has the form $a_{\l}^{\mu}=\frac{c_0}{2}
\delta_{\l\mu}N^2+\beta_{\l\mu}N+\gamma_{\l\mu}$. 
Note that this 
matrix is not Hermitian since our initial basis $\{B_{\l}\}$ is 
orthogonal but not orthonormal, i.e., $\langle B_{\l}|B_{\l}\rangle=
d_{\l}\prod_i\l_i!$. 
Since the interaction is Hermitian, changing the 
basis to orthonormal would render the matrix  $\{a_{\l}^{\mu}\}$ Hermitian.
The matrix $\{a_{\l}^{\mu}\}$ has dimension ${\cal P}(L)$, which is the
 number of 
partitions of $L$.
It had been shown\cite{cl2,15} that the eigenvalue  problem can be reduced to 
${\cal P}(L)-{\cal P}(L-1)-1$ dimensions and recursively solved for the
general interaction up to $L=5$. For $L=6$, the problem reduces to the 
diagonalization of the $3\times 3$ matrix, which can be accomplished using the
$1/N$ expansion. Motivated by this approach, in this 
paper we propose a similar strategy.
In the limit where the \an \ $L$ is much smaller than the number of particles 
$N$  we use perturbation theory in the large-$N$ expansion to calculate 
the interaction energies and derive analytical results.  
In the zeroth order, the standard
perturbation-theory approach gives $A_{\l}=B_{\l}$ for eigenstates
 and $E_{\l}^{(0)}=a_{\l}^{\l}$ for the corresponding energy.
The eigenenergy with the first-order corrections is 
\bee\label{el1}
E_{\l}^{(1)}=a_{\l}^{\l}+\sum_{\mu\neq \l}\frac{a_{\l}^{\mu}a_{\mu}^{\l}}
{a_{\l}^{\l}-a_{\mu}^{\mu}},\eeq
The above expression is applicable if 
 the condition $a_{\l}^{\mu}a_{\mu}^{\l}<<(a_{\l}^{\l}-a_{\mu}^{\mu})^2$ is 
satisfied for all partitions $\mu\neq \l$.
It can be easily checked using relations (\ref{n2}) and (\ref{n3}) that 
$a_{\l}^{\mu}a_{\mu}^{\l}\lsim N$ and $(a_{\l}^{\l}-a_{\mu}^{\mu})^2\sim N^2$.
One finds that the dominant contributions are those 
with $j$ or $l$ equal to zero, in Eq.(\ref{n3}), 
and they produce corrections to 
the energy of order $N^0$.

We can label the exact interaction energies and eigenstates as 
$E_{\l}$ and $A_{\l}$, respectively, such
that in the limit $N\rightarrow\infty$ and finite $L$, the  energy $E_{\l}$ goes
 to $E_{\l}^{(0)}$ and $A_{\l}$ goes to $B_{\l}$.
For an exact eigenstate $A_{\l}(N,L)$, the state $B_1^nA_{\l}(N,L)$ is an 
exact eigenstate $A_{\l'}(N,L+n)$, where $\l'=0^{(n_0-n)}1^{(n_1+n)}
2^{n_2}\cdots$.
According  to  translational invariance, we obtain the 
exact identity for eigenenergies
\bee\label{iden}
E_{0^{(n_0-n)}1^{(n_1+n)}2^{n_2}\cdots}=E_{0^{n_0}1^{n_1}2^{n_2}\cdots}\;.\eeq
Hence, $A_{0^{n_0}1^{n_1}2^{n_2}\cdots}=B_1^{n_1}A_{0^{(n_0+n_1)}
2^{n_2}\cdots}$ and 
$E_{0^{n_0}1^{n_1}2^{n_2}\cdots}=E_{0^{(n_0+n_1)}2^{n_2}\cdots}$.
The part $1^{n_1}$ in the partition $\l$ denotes $n_1$ unit angular momenta 
which can be realized only as the angular momenta due to the center-of-mass 
motion.
Therefore we consider only the eigenstates with partition 
$\l=0^{n_0}2^{n_2}3^{n_3}\cdots l^{n_l}\cdots$, i.e., the states involving 
quadrupoles, octupoles, and higher $l$ poles. 
In this case, we have
\bee\label{E00}
E_{\l}^{(0)}=c_0{n_0(\l)\choose 2}+\sum_{i\geq 2}\alpha_{i0}^{i0}n_i(\l)
n_0(\l)+\sum_{i\geq j\geq 2}\alpha_{ij}^{ij}\frac{2-\delta_{i,j}}{2}
n_i(\l)[n_j(\l)-\delta_{i,j}].\eeq
For special partition $\l=0^{(N-1)}l^1$, we obtain the excitation
energy for a general weak interaction $\epsilon_l=
E_{\l}^{(0)}-E_{0}^{(0)}=N(\alpha^{l0}_{l0}-\alpha^{00}_{00})$. In the case 
of contact interaction it reduces to the 
$\epsilon_l=-c_0N(1-2^{-(l-1)})$\cite{mott}. 
Now, we include corrections
\baa\label{corr}
E_{\l}^{(1)}=E_{\l}^{(0)}&+&
\sum_{i\geq j\geq 2}c_{ij}n_i(\l)[n_j(\l)-\delta_{i,j}][n_{i+j}(\l)+1]\nn
&-&\sum_{i\geq j\geq 1}c_{ij}n_{i+j}(\l)[n_i(\l)+1+\delta_{i,j}][n_j(\l)+1],\ea
where
\bee
c_{ij}=\frac{\alpha_{ij}^{i+j,0}\alpha^{ij}_{i+j,0}\frac{2-\delta_{i,j}}{2}}
{\alpha_{i0}^{i0}+\alpha_{j0}^{j0}-\alpha^{i+j,0}_{i+j,0}-\alpha_{00}^{00}}.\eeq

In the case of repulsive delta interaction,
Eqs.(\ref{E00}) and (\ref{corr}) simplify significantly, because
all coefficients  $c_n$ are zero for $n\neq 0$, so
$\alpha_{ij}^{kl}=2^{(-n)}(2-\delta_{k,l}){n\choose k}c_0,\;n=i+j=k+l$.
From Eq.(\ref{E00}) we easily find the lowest-order energy $E_{\l}^{(0)}$
\bee
E_{\l}^{(0)}=c_0\left\{\frac{N^2}{2}-N\left[\frac{L+2}{4}-\sum_{i\geq 4}
\left(\frac{i}{4}+\frac{1}{2^{i-1}}-1\right)n_i\right]\right\}.\eeq
One can calculate the first-order correction for arbitrary partition, so,
for example,
the eigenenergy for the partition $\l=0^{n_0}2^{n_2}3^{n_3}4^{n_4}$ is
\baa\label{den}
E_{\l}^{(1)}&=&c_0\left\{\frac{N^2}{2}-\frac{N}{8}\left(2L+4-n_4\right)
+\frac{27}{68}n_3(n_3-1)\right.\nn &+&\left.n_4\left(
\frac{81}{52}n_2 +\frac{27}{41}n_3
+\frac{99}{194}n_4+\frac{93}{388}\right)\right\} +{\cal O}(1/N) .\ea
 This result  is in complete agreement with the results obtained in
Refs.\cite{22,cl2}. We see that in the special case of $n_l=0$ for $l\geq 4$,
 the
zeroth-order energy 
$$E_{\l}^{(0)}=c_0\left[\frac{N^2}{2}-\frac{N(L+2)}{4}\right]$$ is
degenerated, but the corrections $\frac{27}{68}n_3(n_3-1)$ remove this
degeneracy if $n_2\geq 2$. Hence, for the repulsive delta interaction the
ground state is unique and defined by $N=n_0+n_2,\;L=2n_2$ or
$N=n_0+n_2+1,\;L=2n_2+3$, depending on $L$ being even or odd.
Therefore, our analysis confirms the conjecture of Smith and Wilkin\cite{13},
in
the limit of large $N$ and finite $L$.

The exact eigenstates for $L\leq 5$ do not depend on details of interaction, 
and can be expanded in the standard basis $B_{\l}$. For example, in the
$L=2$ case we have $B_1^2=B_2+2B_{11}$ and $A_2=\frac{1}{N}B_{11}-
\frac{N-1}{2N}B_2$ and in the $L=3$  case we have
\baa
B_1^3&=&B_3+3B_{21}+6B_{111},\nn
B_1A_2&=&-\frac{N-1}{2N}B_3-\frac{N-3}{2N}B_{21}+\frac{3}{N}B_{111},\nn
A_3&=&\frac{(N-1)(N-2)}{3N^2}\left[B_3-\frac{3}{N-1}B_{21}+\frac{12}{(N-1)(N-2)}
B_{111}\right].\nonumber\ea
It is interesting to consider the probability of the configuration 
$B_{\mu}$ in the exact eigenstate $A_{\l}$, since it gives us a 
physical picture of the excitations. One can easily calculate  the probability
using the formula
\bee\label{prob}
w_{\mu}(A_{\l})=\frac{\langle B_{\mu}|A_{\l}\rangle^2}{\langle B_{\mu}|
B_{\mu}\rangle\langle A_{\l}|A_{\l}\rangle}.\eeq
For the $L=3$ case, the probabilities are given in Table 1. We see that the 
probability of the configuration $B_{1^L}$ in any exact state other than 
$A_{1^L}$ tends to zero, in the limit when $N\rightarrow\infty$ and 
$L$ finite. Of course, this is not a surprise, as we labeled 
the exact states to obey the condition $w_{\mu}(A_{\l})\rightarrow
\delta_{\mu,\l}$ in the large-$N$ limit.

The eigenstates $\e_L=\sum(z_{i_1}-B_1/N)\cdots(z_{i_L}-B_1/N)$, with 
$L\leq N$ are common to all interactions and these are the ground states for 
the repulsive delta interaction\cite{14,15}. 
There is a simple relation between these 
states and the exact eigenstates $A_{\l}$ we have discused up to now. Generally,
$\e_L=A_{2^{L/2}}$ for even $L$ and $\e_L=A_{32^{(L-3)/2}}$ for odd $L$.
For $L<<N$ $\e_L$ is dominated by $\{2^{L/2}\}$ or $\{32^{(L-3)/2}\}$ 
configurations, depending on $L$ being even or odd.

 Of special interest are "vortex" states $\e_{L=N}$. The probability that every 
particle carries a unit of \an \ is easily calculated for low $N$ and is 
given by $w_{11}(\e_2)=1/2$, $w_{1^3}(\e_3)=4/9$, $w_{1^4}(\e_4)=15/32$, 
$w_{1^5}(\e_5)=296/625$. It seems that $w_{1^N}(\e_N)\lsim 1/2$, 
and this is in contrast with the 
naive expectation that all particles contribute with a unit of 
\an \ in the vortex state $\e_{N}$\cite{13}. Note that in this case our 
perturbative approach is not valid, hence the probabilities of 
configurations $\{2^{L/2}\}$ and $\{32^{(L-3)/2}\}$ are small.

In conclusion, we studied a trapped 
Bose-Einstein condensate under rotation in the limit
of weak, translational and rotational invariant two-particle interactions.
We have used the perturbation-theory approach to
calculate the ground-state energy and the excitation spectrum in the
asymptotic limit where the total number of particles $N$ goes to
infinity while keeping the total \an \ $L$ finite. Calculating the
probabilities of  configurations $B_{\mu}$
  in the exact eigenstates $A_{\l}$
gives us a clear view on the physical content of excitations $A_{\l}$.
In addition,  we have briefly discussed
the case of repulsive delta interaction.

Acknowledgment

This work was supported by the Ministry of Science and Technology of the
 Republic of Croatia under Contract No. 00980103.

\narrowtext
\begin{table}
\begin{tabular}{c|c c c}
 &$B_{111}$&$B_{21}$&$B_{3}$ \\ \hline
$B_1^3$&
${\displaystyle\left(1-\frac{1}{N}\right)\left(1-\frac{2}{N}\right)}$
&$ {\displaystyle\frac{3}{N}
\left(1-\frac{1}{N}\right)}$ & ${\displaystyle\frac{1}{N^2}}$ \\
$B_1A_2$ & ${\displaystyle\frac{3}{N}\left(1-\frac{2}{N}\right)}$ & 
${\displaystyle\left(1-
\frac{3}{N}\right)^2}$ & ${\displaystyle\frac{3}{N}\left(1-\frac{1}{N}\right)}$
 \\
$A_3$ &${\displaystyle\frac{4}{N^2}}$ 
& ${\displaystyle\frac{3}{N}\left(1-\frac{2}{N}\right)}$ &
${\displaystyle\left(1-\frac{1}{N}\right)\left(1-\frac{2}{N}\right)}$
\end{tabular}
\vskip0.5pc
\caption{The probabilities of configurations $B_{\mu}$ in exact states $A_{\l}$
for the $L=3$ case.}
\label{table}
\end{table}

\end{document}